# Criterion for transformation of transverse domain wall to vortex or antivortex wall in soft magnetic thin-film nanostripes

Youn-Seok Choi, Sang-Koog Kim,[a] Jun-Young Lee, Myung-Woo Yoo, Konstantin Yu. Guslienko, and Ki-Suk Lee

*Research Center for Spin Dynamics & Spin-Wave Devices, Seoul National University*
*Nanospinics Laboratory, Department of Materials Science and Engineering, Seoul National University, Seoul 151-744, Republic of Korea*

We report on the criterion for the dynamic transformation of the internal structure of moving domain walls (DWs) in soft magnetic thin-film nanostripes above the Walker threshold field, $H_\text{w}$. In order for the process of transformation from transverse wall (TW) to vortex wall (VW) or antivortex wall (AVW) occurs, the edge-soliton core of the TW-type DW should grow sufficiently to the full width at half maximum of the out-of-plane magnetizations of the core area of the stabilized vortex (or antivortex) by moving inward along the transverse (width) direction. Upon completion of the nucleation of the vortex (antivortex) core, the VW (AVW) is stabilized, and then its core accompanies the gyrotropic motion in a potential well (hill) of a given nanostripe. Field strengths exceeding the $H_\text{w}$, which is the onset field of DW velocity breakdown, are not sufficient but necessary conditions for dynamic DW transformation.

[a]Author to whom all correspondence should be addressed. E-mail: sangkoog@snu.ac.kr

Numerous recent studies in the research fields of nanomagnetism and magnetization (**M**) dynamics have focused on magnetic-field or spin-polarized-current-driven motions of domain walls (DWs) in patterned magnetic elements such as thin-film nanostripes,[1-7] owing to the potential applications to solid-state data-storage and logic devices.[8-14] One of longstanding fundamental issues under debate in these research areas is the underlying physics of the breakdown of DW velocity and oscillatory DW motions under magnetic fields exceeding a threshold field known as the Walker field, $H_\text{w}$. This question has been approached by Walker,[15] Thiaville et al.[16] and Nakatani et al.[3] within the context of one-dimensional (1D) DW models. These 1D models can partially explain the linear increase of DW velocity with the field strength in low-field regions, as well as the Walker breakdown behavior in intermediate-field regions.

Alternatively, 2D dynamic soliton models recently developed by Lee et al.,[17] Tretiakov et al.,[18,19] and Guslienko et al.[20] have begun to explain the DW dynamics in the low- and intermediate-field regions more generally, with reference to the dynamic transformation of the internal structure of a moving DW between transverse wall (TW) and antivortex wall (AVW) or vortex wall (VW). According to these 2D models, the dynamic transformation is just the result of the repetitive serial processes of the nucleation (emission), gyrotropic motion (propagation), and annihilation (absorption) of topological solitons such as vortex (V) or antivortex (AV), with the conservation of total topological charges inside a given nanostripe maintained.[17-20] In addition, DW dynamics and its related **M** reversals under higher magnetic fields beyond the Walker breakdown regime can be explained in terms of the nucleation and annihilation of the V-AV topological soliton pairs.[21]

Although such nontrivial DW dynamics in patterned magnetic nanostripes can be understood in the context of the 2D dynamic soliton models and/or 1D models developed thus far, the process of the onset of the dynamic transformation of the TW to the VW (or AVW), along with its criterion, have yet been clarified. In this letter, we present the results of a study on the criterion for dynamic DW transformation, and also suggest a simple way to manipulate the transformation to avoid the reduction (breakdown) of DW velocities above the $H_\text{w}$.

In the present micromagnetic calculations, we used Permalloy (Py: $Ni_{80}Fe_{20}$) thin-film nanostripes of rectangular cross-section, and 10 nm thickness, 200 nm width, and 6.0 μm length, as illustrated in Fig. 1(a). The material parameters chosen were as follows: the saturation magnetization $M_\text{s} = 8.6 \times 10^5$ A/m, the exchange stiffness $A_\text{ex} = 1.3 \times 10^{-11}$ J/m, and zero magnetocrystalline anisotropy. To examine the **M** dynamics of individual cells of 2.5 nm × 2.5 nm × 10 nm dimensions with the Gilbert damping constant of $\alpha = 0.01$, we used the OOMMF code[22] that employs the Landau-Lifshitz-Gilbert equation of **M** motion.[23] As illustrated in Fig. 1(a), we started with a head-to-head TW-type DW (of net **M** in the +y direction) as the initial structure, which DW was positioned at the center of the long axis of a given nanostripe. The TW motion was driven by the $H$ applied in the $+x$ direction. We chose field strengths greater than the $H_w$, which fields are necessary for the transformation of the TW to either the VW or AVW, as known from earlier studies.[17,21,24,25] For example, the dynamic transformation of the internal structure of a moving DW under a static field of $H = 35$ Oe $> H_w$ is illustrated in snapshot images representing the serial changes from TW$_V$ → VW$_{up}$ → TW$_\Lambda$ → VW$_{down}$, and again back to TW$_V$ (where the subscripts indicate the soliton core polarizations)[Fig. 1(b)]. For the cases of $H > H_w$, such DW internal structural changes take place, but different types of structural changes have also been observed, according to the $H$ strength and the dimensions of a nanostripe.[17] For $H < H_w$, however, such transformations never occur, but the TW rather shows a steady movement along nanostripes while its edge-soliton cores are kept in the same relative. Thus, the $H_w$ is the onset field above which the dynamic changes of the internal DW structure occur.

To determine the mechanism of such transformations, we investigated the detailed nucleation process of the V (AV) at the stripe edges, i.e., the stabilization process of the VW (AVW). To do so, in place of static fields, we applied step-pulse fields, as shown in the inset of Fig. 1(a), with different duration times $\Delta t_H$ and with equal field strength $H = 35$ Oe. Here, we chose a specific transformation, TW$_V$ → VW$_{up}$, which is a part of the internal structural changes in a unit period: TW$_V$ → VW$_{up}$ → TW$_\Lambda$ → VW$_{down}$ [Fig. 1(b)], namely the type (ii) noted in Ref.17. The VW-type DW has a potential *well* along the width direction, and hence, when a moving TW turns into a VW, the forward motion of the initial TW becomes the backward motion of the VW-type DW. Accordingly, the onset of the DW transformation can be determined by examining the tuning point of the direction of the DW motion.

The trajectories of the up-edge soliton cores in motions driven by the step-pulse fields of $\Delta t_H$ = 0.600, 0.610, 0.614, 0.700, and 0.800 ns with $H$ = 35 Oe > $H_w$, are shown in Fig. 2(a). For the shorter times, $\Delta t_H$ = 0.600, 0.610, and 0.614 ns, the TW$_V$ generally maintained its original **M** configuration and moved along the long axis of the nanostripe. Even after the pulse fields were turned off (those times were marked with small lines on the trajectories), the edge-soliton core of the TW moved inward slightly, but immediately returned to the edge, as shown in those trajectories. By contrast, for the longer pulse durations, $\Delta t_H$ = 0.700 and 0.800 ns, the TW changed irreversibly its **M** configuration to a VW-type DW, after which it started to move backward in the longitudinal (*x*-axis) direction and move farther in the transverse direction, as shown by serial snapshot images in Fig. 2(b). The backward motion of the vortex core shows oscillatory motions in the transverse direction, which may be due to the interaction between the nucleated vortex core and the edge-soliton core with its polarization reversal initially placed at the down-stripe edge [see images with indicated number ④ and ⑤ in Fig. 2(b)].

Such contrasting DW motions, driven by the pulse fields with longer and shorter durations with the equal strength $H$ = 35 Oe > $H_w$, cannot be explained in the context of the so-far-developed 1D models. To understand such contrasting behaviors more quantitatively, we separately plotted the position deviations, $\Delta y$, of the vortex core in the transverse direction from the up-stripe edge [Fig. 3(a)], along with its longitudinal displacements [Fig. 3(b)], as functions of $t \gg \Delta t_\mathrm{H}$. For the case in which the edge-nucleated vortex core moves up to 12.5 nm from the upper stripe-edge, the core immediately returns to the edge, and consequently the TW moves farther in the +x direction. On the contrary, for the case in which the vortex core [shown by snapshot images for $\Delta t_\mathrm{H}$ = 0.800 ns in Fig. 2(b)] moves 15 nm in the transverse direction, the backward motion of the DW along the longitudinal direction initiates, as in the characteristic gyrotropic motion of a vortex core in a given potential well.[26]

It is certain that there is a critical transverse deviation of the vortex core position (maximum $m_z = M_z/M_s$) from the stripe edge, $\Delta y_\mathrm{cri}$, that determines the onset of the backward motion of the DW in the longitudinal direction, which reflects the transformation of a TW- to VW-type wall. The value of $\Delta y_\mathrm{cri}$ is found to be larger than 12.5 nm and less than 15.0 nm from our simulation results, as shown in Fig. 3(c). The $\Delta y_\mathrm{cri}$ value between 12.5 and 15.0 nm is close to $\Delta w_\mathrm{FWHM}$, the full width at half maximum (FWHM) of the $m_z$ of the stabilized vortex core, indicating that the VW (or AVW) can be stabilized when its core position moves inward along the transverse direction sufficiently to $\Delta w_\mathrm{FWHM}$ ~14 nm, which value is estimated from Fig. 3(c).

The Walker field is known, as yet, to be the onset field above which the DW velocity breaks down, leading to oscillatory backward and forward DW motions in nanostripes. However, on the basis of the present results, the fields exceeding the $H_w$ are not the sufficient, but rather the necessary condition for dynamic changes of the DW internal structure from the TW to the VW (or AVW). For DW dynamic transformations, the true sufficient criterion required for stabilizing the core of the V (or AV) is found to be $\Delta y_{cri}$ ~14 nm. This also is a reflection of the fact that 2D dynamic soliton models of DW dynamics are more generally appropriate than 1D models for interpreting and understanding the internal structural changes of moving DWs above the $H_w$.

To confirm the applicability of 2D soliton models for the interpretations of the above results, we also compared the simulation results of the trajectories of the backward motion of a vortex core after it was nucleated and stabilized, with the calculation results of an analytical equation of soliton-core trajectories obtained based on the 2D dynamic soliton model.[20] After the nucleation of the vortex core, its motion was driven after pulse fields were off, i.e, by $H = 0$. The trajectories of the vortex-core backward motion can thus be obtained by inputting $H = 0$ into the parabolic trajectory equation for a certain field $H > H_w$, described in Ref. 20, such that $Y(t) - Y(0) = -(1/C)[X(t) - X(0)]$. The $X(t)$ and $Y(t)$ correspond to the $x$ and $y$ components of the soliton core position. We chose the value of $[X(0), Y(0)]$ as the core position at which the VW core starts to move backward in the $x$ direction. The trajectory slope, $[Y(t)-Y(0)]/[X(t)-X(0)]$ is simply $-1/C$. For the geometry and dimensions of the given nanostripe, the value of $C$ was estimated to be -21.707 (Ref. 20), which is in quantitative agreement with the simulation results, as shown in Fig. 4.

The above results pertaining to the criterion for the dynamical nucleation of a vortex core near the stripe edge are important from an application point of view. They show that periodic pulse fields can be implemented to manipulate the internal structure of a moving DW, in particular, in order to avoid the DW transformation that yields the velocity breakdown above the $H_w$. Figure 5 shows a comparison of the overall DW motions driven by a static field of $H$ = 35 Oe > $H_w$, and a periodic pulse field ($H$ = 35 Oe) of duration time $\Delta t_H$ = 0.6 ns and accompanying relaxation time $\Delta t_R$ = 2.5 ns. For the case of the static field in Fig. 5(a), the DW internal structure changes, as already shown in Fig. 1(b), thus the different DW structures show their characteristic motions, yielding the oscillatory behavior of the longitudinal displacement and velocity. Contrastingly, the periodic pulse field does not allow for the transformation of the TW to the VW [Fig. 5(b)], and thus the average velocity of the TW is 350 m/s, which is 4.7 times greater than the average velocity, 75 m/s for the oscillatory DW motion driven by the static field. It is worthwhile noting that the application of such periodic pulse fields offers an efficient way to prohibit the breakdown (reduction) of DW velocities above the $H_w$, without using the edge roughness of a given nanostripe,[3] an underlayer of strong perpendicular magnetization,[25] and the application of perpendicular magnetic fields.[27]

In conclusion, the present study on the onset of the nucleation process of a vortex core driven by step-pulse fields offers a physical explanation of the criterion for the domain wall dynamic transformations observed experimentally, as well as a simple way to control the domain wall velocity breakdown typically observed above the Walker field.

## Acknowledgement

This work was supported by Creative Research Initiatives (ReC-SDSW) of MEST/KOSEF.

**Figure captions**

Fig. 1. (Color online) (a) Thin-film Py nanostripe of indicated dimensions, along with initial **M** configuration of TW-type DW placed at middle of long axis. The in-plane orientations of the local **M**s are represented by the colors and streamlines. The inset shows a step-pulse field with duration time $\Delta t_H$. (b) Plane-view snapshot images of dynamic transformation of moving DWs driven by a static field $H= 35$ Oe. The small arrows and black and white colors indicate the local in-plane **M** orientations.

Fig. 2. (Color online) (a) Trajectories of moving edge-soliton cores placed initially at the upper edge in nanostripe under applied step-pulse fields with indicated values of $\Delta t_H$. The small lines on the trajectories correspond to the times at which the pulse fields were turned off in each case. (b) Plane-view snapshot images of those temporal evolutions taken at times marked by small arrows with numbers on the trajectory for $\Delta t_H =0.800$ ns.

Fig. 3. (a) Transverse displacement from upper edge ($\Delta y$) and (b) longitudinal ($x$) displacement of the position of the upper edge-soliton core for each case of $\Delta t_H$ shown in Fig. 2(a). (c) $m_z = M_z/M_s$ profiles along the transverse direction across the upper edge-soliton core moving under different values of $\Delta t_H$. The dotted horizontal line corresponds to $m_z = 1/2$.

Fig. 4. (Color online) Trajectories of the backward motion of a vortex core after it was nucleated and stabilized (open circles), initially driven by step-pulse fields of $\Delta t_H$ = 0.700 and 0.800 ns, and the calculation results of an analytical equation of soliton-core trajectories obtained based on the 2D dynamic soliton model (solid line). The backward motions were examined for much longer times than $\Delta t_H$. The solid symbols correspond to the trajectories of the TW-type DW before its transformation to the VW-type DW.

Fig. 5. (Color online) Longitudinal (*x*) displacements and velocities of DW motions driven by static field of *H* = 35 Oe in (a) and by periodic step-pulse field with the same field strength and with $\Delta t_H$ = 0.600 ns and $\Delta t_R$ =2.5 ns in (b).

Fig. 1

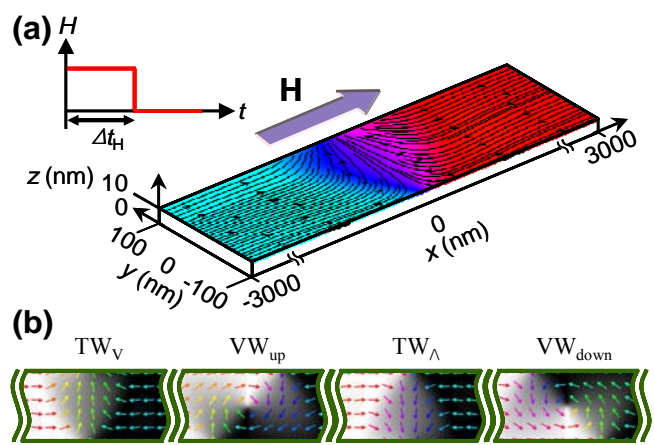

Fig. 2.

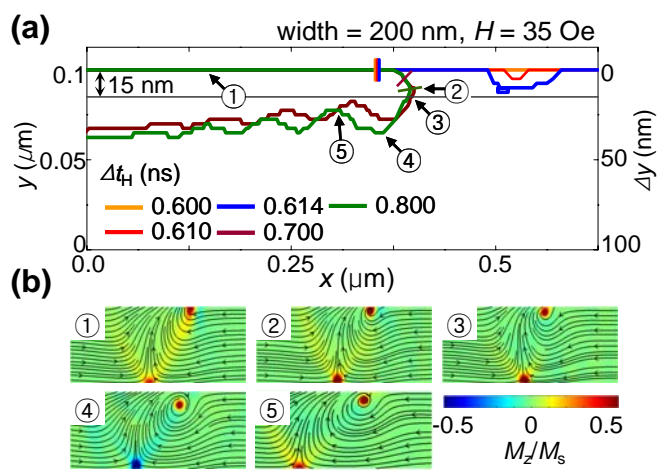

Fig. 3

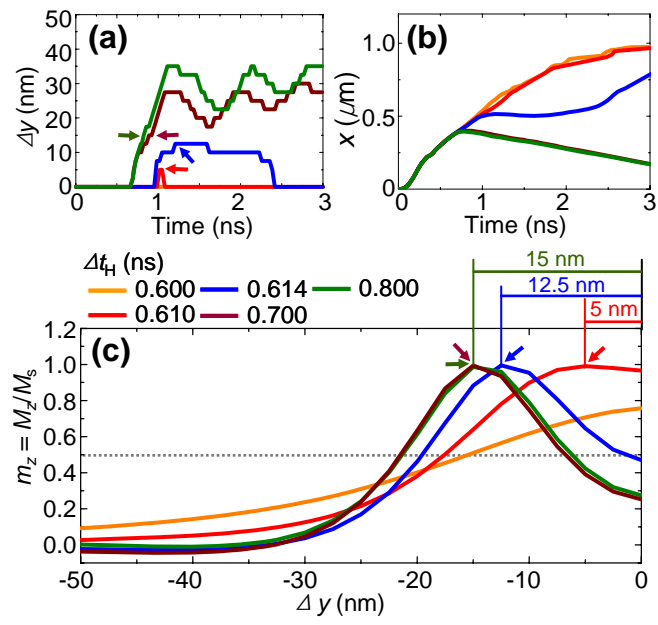

Fig. 4.

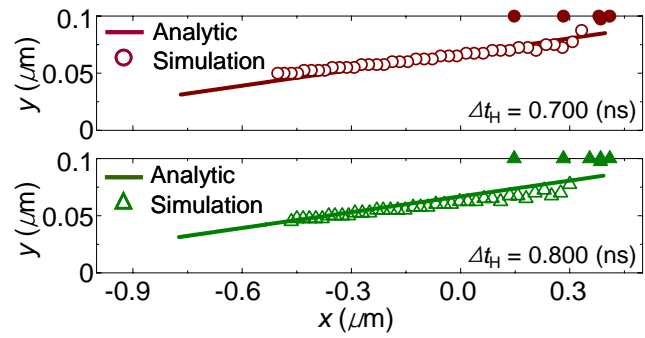

Fig. 5.

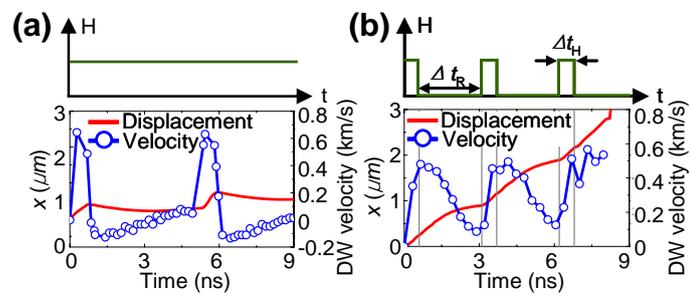